\documentclass[preprint,prd,noshowpacs,nofootinbib]{revtex4}
\usepackage{amssymb}
\usepackage{amsmath}
\usepackage{amsfonts}
\usepackage{graphicx, float}

\begin{document}

\title{Baryogenesis via Hawking-like Radiation in the FRW Space-time}

\author{Sujoy K. Modak}
\email{sujoy.kumar@correo.nucleares.unam.mx}
\affiliation{Instituto de Ciencias Nucleares, Universidad Nacional Aut{\'o}noma de M{\'e}xico,
Apartado Postal 70-543, Distrito Federal, 04510, M{\'e}xico}
\author{Douglas Singleton}
\email{dougs@csufresno.edu}
\affiliation{Department of Physics, California State University, Fresno, CA 93740-8031 USA}
\affiliation{Instituto de Ciencias Nucleares, Universidad Nacional Aut{\'o}noma de M{\'e}xico,
Apartado Postal 70-543, Distrito Federal, 04510, M{\'e}xico}

\date{\today}

\begin{abstract}
We present a phenomenological model for baryogenesis based on particle creation in the Friedman-Robertson-Walker (FRW) spacetime. This study is a continuation of our proposal that Hawking-like radiation in FRW space-time explains several physical aspects of the early Universe including inflation. In this model we study a coupling between the FRW space-time, in the form of the derivative of the Ricci scalar, and the $B-L$ current, $J^{\mu} _{B-L}$, which leads to a different chemical potential between baryons and anti-baryons resulting in an excess of baryons over anti-baryons with the right order of magnitude. In this model the generation of baryon asymmetry, in principle, occurs over the entire history of the Universe starting from the beginning of the radiation phase. However, in practice, almost the entire contribution to the baryon asymmetry only comes from the very beginning of the Universe and is negligible thereafter. There is a free parameter in our model which can be interpreted as defining the boundary between the unknown quantum gravity regime and the inflation/baryogenesis regime covered by our model. When this parameter is adjusted to give the observed value of baryon asymmetry we get a higher than usual energy scale for our inflation model which however may be in line with the GUT scale for inflation in view of the BICEP2 and Planck results. In addition our model provides the correct temperature for the CMB photons at the time of decoupling.          
\end{abstract}

\maketitle

\section{Introduction}

One of the open questions in cosmology is the generation of a Universe filled with more baryons than
anti-baryons, as required by observation. The quantity which captures this predominance of baryons over 
anti-baryons is the parameter $\eta$ defined as
\begin{equation}
\label{eta}
\eta = \frac{n_B - n_{\bar B}}{s} = \frac{n_B}{s} ~,
\end{equation}
where $n_B$ is the volume density of the number of baryons and $s$ is the volume entropy density. Current measurements \cite{eta} give $5.1 \times 10^{-10} < \eta < 6.5 \times 10^{-10}$ for which we will use the approximation $\eta \approx 6 \times 10^{-10}$  throughout the paper. Sakharov \cite{Sakharov} was the first to point out three general conditions apparently needed to generate a baryon asymmetry: (i) violation of baryon number; (ii) violation of charge ($C$) and charge-parity ($CP$) symmetry; (iii) departure from thermal equilibrium. In \cite{steinhardt} a mechanism was given whereby one could generate the required baryon asymmetry while being in thermal equilibrium {\it i.e.} one could generate a baryon asymmetry with only the first two of Sakharov's conditions. The idea of \cite{steinhardt} was to introduce a quantum gravity inspired coupling between the space-time and the baryon minus lepton current 
$J^{\mu} _{B-L}$ whose action has the form 
\begin{equation}
\label{rj}
\frac{\hbar ^3}{M^2 _* c} \int d^4 x \sqrt{-g} (\partial _\mu {\cal R})  J^{\mu} _{B-L} ~,
\end{equation}  
where ${\cal R}$ is the Ricci scalar of the space-time and $M_*$ is some mass scale. Following \cite{steinhardt} we will take this mass scale to be the reduced Planck mass 
$M_* \approx ((\hbar c)/(8 \pi G))^{1/2} \approx 2.4 \times 10 ^{18} \rm ~ {GeV}$. An interaction such as
\eqref{rj} is excepted to emerge from quantum gravity. From \eqref{rj} one can define a chemical
potential, for certain species of particles carrying $B-L$ charge as
\begin{equation}
\label{chem-pot}
\mu _i = \frac{q_i \hbar ^3}{c^2} \frac{{\cal {\dot R}}}{M^2 _*}~,
\end{equation}
where $i$ is a particle species index which labels the particles carrying a $B-L$ charge of $q_i$ and  
$\partial _\mu {\cal R} \rightarrow  {\cal {\dot R}} = \frac{d{\cal R}}{dt}$ is the time derivative of the Ricci 
scalar with the space-time taken to be varying temporally but not spatially.

Since baryons and anti-baryons have different sign charges, $q_i$, the chemical potentials will be different for them. In the presence of a thermal bath of temperature $T$ the number difference between baryon and anti-baryons per volume is \cite{hook}
\begin{equation}
\label{delta-N}
\frac{d(\Delta N _{B-L})}{d(Area) (c dt)} = \int \frac{d^3p}{(2 \pi \hbar )^3} \frac{1}{e^{(p c  - \mu)/ k_B T} +1}
- \int \frac{d^3p}{(2 \pi \hbar)^3} \frac{1}{e^{(p c + \mu)/ k_B  T} +1} ~,
\end{equation}
Note the differential volume in \eqref{delta-N} is the product of an area differential multiplied by 
a time differential -- $d(Area) (c dt)$ -- with the $dt$ being a length after multiplication
by $c$. In \eqref{delta-N} we have only considered fermions since we are using the Pauli-Dirac distribution 
$\frac{1}{e^{(p c \pm m)/ k_B T} +1}$ modified in this case by the presence of the chemical potential $\mu$. The reason for this is that in the Standard Model only fermions carry baryon and lepton number. Since the chemical potentials for baryons and anti-baryons are different one gets a non-zero value for $\frac{d(\Delta N _{B-L})}{d(Area) (c dt)}$. Carrying out the integral over $p$ one finds that the first and second integrals in \eqref{delta-N} are proportional to $T^3 Li_3 (-e^{- \mu / k_B T})$  and 
$T^3 Li_3 (-e^{\mu / k_B T})$ respectively, where $Li_3 (x)$ are third order Polylog functions of $x$. Taking the limit 
$\mu \ll k_B T$ and putting back all the numerical factors one finds that \eqref{delta-N} becomes
\begin{equation}
\label{delta-N2}
\frac{d(\Delta N _{B-L})}{d(Area)(c dt)} \approx \frac{\mu ^3}{6 \pi^2 (\hbar c )^3 } + 
\frac{\mu k_B ^2 T^2}{6 (\hbar c )^3 } \ \approx \frac{\mu k_B ^2 T^2}{6 (\hbar c )^3}~,
\end{equation}
where in the last step we have dropped $\mu^3$ relative to $\mu T^2$ using $\mu \ll T$. The results in \eqref{delta-N} and \eqref{delta-N2} are for one type of fermion and for one degree of freedom. To take into account different types of fermions with $B-L$ charges given by $q_i$ and degrees of freedom $g_i$, one should sum over different types of fermions with different degrees of freedom, so that
\begin{equation}
\label{delta-N3}
\frac{d(\Delta N _{B-L})}{d(Area) (c dt)} = \sum _i g_i q_i \frac{\mu_i k_B ^2 T^2}{6 (\hbar c )^3} =
\frac{k_B ^2 T^2}{6 M^2 _* c^5} \sum _i g_i q_i ^2 {\cal {\dot R}} ~,
\end{equation}
where we have used \eqref{chem-pot} in arriving at the last result in \eqref{delta-N3}. 

The sum in \eqref{delta-N3} goes only over the baryons since we assume that after the baryons and anti-baryons have annihilated what we are left with is the small excess of baryons. This is in accord with what is done with $\eta$ in \eqref{eta} where the final expression is written in terms of the baryon number density $n_B$. A final step that we can perform is to integrate over the $Area$ in \eqref{delta-N3} to give
\begin{equation}
\label{delta-N4}
\frac{d(\Delta N _{B-L})}{dt} = \frac{k_B ^2 T^2 \times Area}{6 M^2 _* c^4} \sum _i g_i q_i ^2 {\cal {\dot R}} ~.
\end{equation}
At this point what is left to compute the above quantity is to fix $T$, $Area$, and ${\cal {\dot R}}$ which is possible once we specify the background space-time. Once this is done we can carry out the time integration in \eqref{delta-N4} to get $\Delta N_{B-L}$ and from this we shall be able to find the ratio $\eta$ in \eqref{eta}.

\section{Inflation via FRW Hawking-like radiation}

Since our model of baryogensis is closely connected with the model of inflation driven by Hawking-like radiation in FRW spacetime, as proposed in \cite{modak1, modak2}, in this section we review these works. The idea that particle creation in a given space-time (a standard example of which is Hawking radiation from a black hole) can drive  inflation has been investigated by various researchers. The earliest example we found was that of Prigogine and co-workers \cite{prigo}. More recent and detailed studies on the effect of particle creation on cosmological evolution can be found in \cite{lima1, lima2, lima3, lima4, lima5, sing, gao, naut}. 

In the presence of generic particle creation the usual FRW equations are modified to take the following form \cite{lima4}
\begin{eqnarray}
3\frac{\dot a^2}{a^2} +3 \frac{k c^2}{a^2} &=& \frac{8\pi G \rho}{c^2} ~\label{feq1} \\
2\frac{\ddot{a}}{a}+\frac{\dot a^2}{a^2}+\frac{kc^2}{a^2} &=& -\frac{8\pi G}{c^2}(p-p_c)~\label{feq2} \\
\frac{\dot{n}}{n}+3\frac{\dot a}{a} &=& \frac{\psi}{n}~.\label{feq3}
\end{eqnarray}
The first two equations \eqref{feq1} and \eqref{feq2} are the standard Friedman equations except that the second equation has an additional term $p_c$ which represents a pressure due to the particle creation. We will shortly define $p_c$. The third equation is new and it represents the relationship among the number density $n$, time rate of change of the particle number density, ${\dot n}$, the space-time scale factor $a$ and the creation rate $\psi$. The creation pressure term $p_c$ is given in terms of the creation rate, $\psi$, by 
\begin{eqnarray}
p_c=\frac{\rho + p}{3 n H}\psi ~, 
\label{crp}
\end{eqnarray}
where $H = \frac{{\dot a}}{a}$ is the usual Hubble parameter. In \cite{modak1, modak2} we studied the above scenario in the case of FRW space-time. First, FRW space-time has a horizon so one can calculate a Hawking-like temperature for it \cite{cai, cai1, zhu} which is approximately given by 
\begin{equation}
T \approx \frac{\hbar H }{2\pi k_B} ~ \label{frw-temp} ~.
\end{equation}
Note this temperature is an approximation which assumes that the spatial curvature of FRW is zero ({\it i.e.} $k=0$ in \eqref{feq1} \eqref{feq2}) and also that the surface gravity, $\kappa$, is well approximated by $\kappa \sim H$. A full discussion and justification of these approximations can be found in \cite{modak1, modak2}. 

The proposed inflation mechanism of \cite{modak1, modak2} is based on a treatment that the net effect of Hawking-like radiation from the apparent horizon of FRW spacetime is an effective power gain in the Universe, given by the Stephan-Boltzmann (S-B) radiation law
\begin{equation}
P=+\frac{dQ}{dt}=\sigma A_H T^4, \label{sb}
\end{equation}
where $\sigma=\frac{\pi^2 k_B^4}{60\hbar^3 c^2}$ is known as the S-B constant, $A_H $ is the area of the apparent horizon. This together with the differential form of the first law of thermodynamics
\begin{eqnarray}
\frac{dQ}{dt}=\frac{d}{dt}(\rho V)+p \frac{dV}{dt} ~,
\label{flaw}
\end{eqnarray}
implies that in presence of particle creation the left hand side of \eqref{flaw} is actually non-zero which in turn implies that the Universe is actually an {\it open}, {\it adiabatic} system \cite{prigo}. As explained in \cite{modak1, modak2} this modification is in line with the fact that the observable Universe has an enormous entropy. The interesting fact is that with this set-up one obtains the modified time development equation for the energy density $\rho (t)$
\begin{equation}
\frac{\dot{\rho}}{\rho}+3(1+\omega)\frac{\dot a}{a} = 3 \omega_c (t) \frac{\dot a}{a} ~\label{neq1}
\end{equation} 
which is completely compatible with the modified Friedman equations \eqref{feq1}, \eqref{feq2} and \eqref{feq3} with an identification of particle creation rate \cite{modak1, modak2}
\begin{equation}
\label{crph1}
\psi_{FRW} (t)=\frac{3 n H \omega_c(t)}{(1+\omega)} .
\end{equation}
Here $n$ is again the number density of particles, $\omega = p/ \rho $ is the standard equation of state  parameter ({\it i.e.} the ratio of pressure to energy density of the matter/radiation source for the space-time),  and $\omega_c(t)$ is the equation of state parameter associated with the creation pressure, $p_c$. It is defined as $\omega_c = p_c / \rho$ with $\rho$ being the same energy density used in defining $\omega$ -- which implies particle creation gives an effective pressure, but not an effective energy density. In \cite{modak1, modak2} it was also shown that this equation of state parameter associated with the creation pressure was
\begin{equation}
\label{omega-c} 
\omega_c (t) = \alpha \rho (t) ~~, {\rm where}  ~~\alpha = \frac{\hbar G^2}{45 c^7}= 
4.8 \times 10^{-116} (J/m^3)^{-1} ~.
\end{equation} 
In the absence of the postulated FRW particle creation the right hand side of \eqref{neq1} would be zero giving back the standard conservation law as found by using the Einstein-Friedman equation.
  
The solution of the differential equation \eqref{neq1} for the energy density $\rho (t)$ is found to be
\begin{equation}
\rho=\frac{D_0 a^{-3(1+\omega)}}{1+(\frac{\alpha D_0 }{1+\omega})a^{-3(1+\omega)}} \rightarrow 
 \frac{D_0}{a^4+\frac{3\alpha D_0 }{4}}~.\label{nend}
\end{equation}
For deriving the last expression we have assumed the equation of state of radiation {\it i.e.} $\omega = 1/3$, since after the inflationary stage the Universe should transit to a radiation dominated stage. Note that from \eqref{nend} one can see that when $\frac{3\alpha D_0}{4}$ term in the denominator dominates we have $\rho \approx const.$ and therefore an approximately de Sitter like phase ({\it i.e.} inflation), while when the $a^4$ term in the denominator dominates one has a radiation dominated stage. The integration constant $D_0$ is set as $D_0 \approx 10^{91} \frac{J}{m^3}$  from the requirement of matching with the late time energy density of radiation \cite{modak1, modak2}. Now using this energy density from \eqref{nend} into the Friedman equations, recalling that we have $k=0$, using the equation of state parameter $\omega =1/3$ to eliminate $p$ in favor of $\rho$, and taking into account the creation pressure $p_c (t)$ from \eqref{crp} one can integrate to get an expression for 
$a(t)$ as \cite{modak1, modak2}
\begin{equation}
\sqrt{\alpha D_0 + \frac{4}{3} a^{4} }+\sqrt{\alpha D_0} 
\ln \left[\frac{a^2}{2\sqrt{3} \left(\sqrt{\alpha D_0}+ 
\sqrt{\alpha D_0+\frac{4}{3} a^4 }\right)}\right] = \frac{8}{3}\sqrt{\frac{2\pi G D}{c^2}}t -(K_0-1) 
\sqrt{\alpha D_0} ~.
\label{soln}
\end{equation}
Here $K_0$ is an integration constant {\it which sets the time when the FRW Hawking-like radiation driven inflation begins} \cite{modak1, modak2}. To get a better picture of the behavior of $a(t)$ from \eqref{soln} we take the limits: (i) $a^4 \ll \frac{3\alpha D_0}{4}$ in \eqref{nend}; (ii) $a^4 \gg \frac{3\alpha D_0}{4}$ in \eqref{nend}. For the limit (i) $\rho \approx const.$ one finds the solution of 
\eqref{soln} to be
\begin{equation}
a(t) = 2 (3\alpha D_0)^{\frac{1}{4}} \exp \left[ {\sqrt{\frac{32\pi G}{9c^2\alpha}} ~ t}  - 
\frac{K_0}{2} \right] ~,
\label{inflation-era}
\end{equation}
{\it i.e.} one finds a de Sitter-like exponential expansion (as is expected for the inflationary era)
with a Hubble parameter given by 
\begin{equation}
\label{H}
H_{(dS)} = \frac{\dot a}{a} = \sqrt{\frac{32\pi G}{9c^2\alpha}} \approx 10^{45}~s^{-1}.
\end{equation}
Whereas, in the limit (ii) ( {\it i.e.} $a^4 \gg \frac{3\alpha D_0}{4}$) $\rho \approx D_0/a^4$ and one finds the solution in 
\eqref{soln} becomes
\begin{equation}
a(t) \approx \left( \frac{32 \pi G D_0}{3 c^2} \right)^{1/4}  t^{1/2} ~, \label{pl}
\end{equation}
{\it i.e.} one finds radiation dominated expansion. Again the constant $D_0 \approx 10^{91} \frac{J}{m^3}$ is set by the requirement that the solution in \eqref{pl} gives the correct current energy density of radiation \cite{modak1} \cite{modak2}. 

As discussed in \cite{modak2}, an interesting feature of this model for inflation is that although the duration of inflation is fixed by the Hubble parameter to be $\Delta t \approx 6 \times 10 ^{-44}~s$, the beginning of inflation is open due to the unknown value of the coefficient $K_0$ -- there is an exponential suppression in the scale factor $a(t)$ due to $K_0$ and effective inflation takes place when $H_{(dS)} t$ surpasses this value. Fixing this constant will be important later when we get to the main goal of this paper which is a discussion of baryogenesis in this model. As an important comment, we should mention that the actual energy scale of inflation is still an open issue. The recent observations of the BICEP2 experiment \cite{bicep2} have suggested a Grand Unified Theory (GUT) energy scale inflation of about $10^{16}$ GeV  which is close to the Planck scale. Also there exist other models, such as, \cite{lqg} where inflation occurs exactly at the Planck scale. Later on we shall argue the beginning of inflation can be fixed in our model by using the observed value of $\eta$ in \eqref{eta} to set $K_0$.

With the basic features of FRW Hawking-like radiation reviewed in this section we now turn to the generation of baryon asymmetric in this model.

\section{Generation of Baryon Asymmetry}

With the background of the previous two sections we now enter the main task of this work -- to find out the details of how baryogenesis works in the FRW Hawking-like radiation driven inflation model of \cite{modak1, modak2}. To this end we return to \eqref{delta-N4}. We first note that we can write the area as the horizon area of the FRW space-time 
$Area = 4 \pi r_{FRW} ^2 = \frac{4 \pi c^2}{H^2}$ since $r_{FRW} = \frac{c}{H}$  (using the assumption that $k=0$) and $T$ in \eqref{delta-N4} by \eqref{frw-temp}. The time derivative of the Ricci scalar (${\dot{\cal R}}$) for FRW space-time is given by 
\cite{steinhardt} \cite{hook} 
\begin{equation}
\label{r-dot}
\dot{\cal R} = - 9 (1 - 3 \omega) (1+\omega) \frac{H^3}{c^2} ~.
\end{equation}

Putting these three quantities into \eqref{delta-N4} yields 
\begin{equation}
\label{delta-N5}
\frac{d(\Delta N _{B-L})}{dt} = - \frac{3 \hbar ^2}{2 \pi M_* ^2 c^4} (1+\omega) (1- 3 \omega)
H^3 \sum _i g_i q_i ^2 ~.
\end{equation}

The first point to note about \eqref{delta-N5} is that to lowest order $\frac{d(\Delta N _{B-L})}{dt} =0$, both during the inflationary de Sitter phase when $\omega = -1$ and as well during the radiation dominated phase $\omega = 1/3$. This is valid classically, however at the one-loop level in the Standard Model one finds \cite{qcd} that 
\begin{equation}
\label{1-3w}
1- 3 \omega = \frac{5}{6 \pi ^2} \frac{g^4}{(4 \pi \hbar c)^2} 
\frac{(N_c + \frac{5}{4} N_f)(\frac{11}{3} N_c - \frac{2}{3}N_f)}{2+ \frac{7}{2}[N_c N_f/(N_c ^2 -1)]} ~,
\end{equation}
where $N_c =3$  is the number of colors, $N_f = 6$ is the number of flavors and $g^2 / (4 \pi \hbar c) \approx 0.1$ is the $SU(3)$ fine structure constant. With these parameters one finds that $1-3\omega \approx 10^{-2} - 10^{-3}$. In this way one can write down definite values for the $\omega$ dependent terms in \eqref{delta-N5}. We also note that while $1+\omega =0$ during the inflation stage, that $1- 3 \omega$ is small but non-zero. This implies the baryon asymmetry in our mechanism must be generated during the radiation dominated phase immediately after inflation. To integrate \eqref{delta-N5} we need to know the time dependence of the Hubble parameter during baryogenesis which we now know must coincide with the onset of the radiation domination stage of the Universe. Thus, $a(t) \propto t^{1/2}$, so that $H = \frac{{\dot a}}{a} =\frac{1}{2t}$. Note that there is a term in \eqref{delta-N5} which includes the sum over the degrees of freedom $g_i$ and the $B-L$ charges squared, $q_i ^2$. For the Standard Model its value is known --  $\sum _i g_i q_i ^2 = 13$. Using these in \eqref{delta-N5} (we consider $1+\omega \approx 4/3$ and $1-3\omega \approx 5.0 \times 10^{-3}$) we find 
\begin{equation}
\label{delta-N6}
\frac{d(\Delta N _{B-L})}{dt} = - \frac{0.13 }{8 \pi M_* ^2 c^4}\frac{\hbar ^2}{t^3} ~.
\end{equation}
Now we just need to time integrate \eqref{delta-N6} with definite limits to obtain $\Delta N _{B-L}$, {\it i.e.}, the total number of baryon excess generated by this mechanism. 

As we noted, in this model, baryogenesis begins with the onset of the radiation dominated phase (since $1 + \omega =0$ during inflation there is no baryon excess generated in this stage). Thus we should start our integration from $t= t_{rad} ^*$ (the beginning time of the radiation domination). Also, since the Hawking-like radiation in FRW space-time should always be there as long as there is a horizon, the upper limit of the integral could be taken as $t=t_{present} \approx \infty$. However for all practical purposes the particle creation effect does not last for any significant time beyond $t= t_{rad} ^*$. For example if we take the upper limit to be $10 \times t_{rad} ^*$ rather than $t=t_{present} \approx \infty$ there will only be a 
1\% difference between taking the upper limit as $t=\infty$ versus $t= 10 t_{rad} ^*$. Because of this feature 
({\it i.e.} that most of the particle creation comes from a very short range of time after $t= t_{rad} ^*$) we will take the upper limit in the integration as $t=\infty$ since there is only a very small difference between taking this as the upper limit in the integration versus taking the upper limit as $t= 10 t_{rad} ^*$ or $t= 100 t_{rad} ^*$ for example. 

At this point we only need to provide a value for $t_{rad}^*$ to get a specific number after performing the integral 
\eqref{delta-N6}. In the following section we show that this value is directly related with the constant $K_0$ in 
\eqref{inflation-era}.

\section{Observed value of baryogenesis and the constant $K_0$}

As discussed in Section 2, in our model of inflation there is a free parameter,$K_0$, that fixes the beginning of inflation. Once this beginning time is fixed the end of inflation, or equivalently, the onset of radiation domination is already fixed since the duration of inflation is fixed in our model as $\Delta t \approx 10 ^{-44} - 10^{-43} {\rm s}$. Therefore, $t_{rad} ^*$ is related to the integration constant $K_0$ in \eqref{inflation-era} {\it i.e.} fixing $K_0$ fixes $t_{rad} ^*$. 

With such a crucial role played by $K_{0}$ one might be interested to have a physical explanation behind this. We first note that since $K_0$ fixes the onset of inflation driven by Hawking-like radiation as one goes from higher to lower energy scales, it also determines the reverse, {\it i.e.}, the switching off the Hawking effect in the reverse direction. On the other hand generically one expects that at some very high energy scale quantum gravity would take over from the semi-classical Hawking effect making the latter invalid. In this respect, $K_0$ can be thought of as a parametrization of our {\it ignorance of quantum gravity regime} since before inflation one is certainly in a regime where quantum gravity is important but the exact time/energy scale of the transition between the quantum gravity regime and the semi-classical regime cannot be determined by semi-classical means {\footnote{This point of ``graceful entrance'' to inflation from teh quantum gravity regime is discussed more elaborately in \cite{modak1, modak2} where it is postulated that in the quantum gravity regime the FRW, Hawking-like radiation -- and our version of inflation -- turns off.}}. Interestingly, in the following we shall show that, for FRW space-time $K_0$ {\it can be fixed} by demanding that we generate the observed value of baryon to entropy density (\ref{eta}).  

First, integrating \eqref{delta-N6} leads to
\begin{equation}
\label{N1}
N _{B-L} = - \frac{0.13 }{8 \pi M_* ^2 c^4} \int _{t_{rad} ^*} ^ \infty \frac{\hbar ^2}{t^3} ~dt =
- \frac{0.13 \hbar ^2}{16 \pi M_* ^2 c^4 (t_{rad} ^*)^2}~.
\end{equation}
In principle baryons are generated via this mechanism even up to the present time which is the reason we
take the upper limit of integration as $t=\infty$ -- there is effectively no difference between taking the 
upper limit as $\infty$ versus taking the upper limit as the present time. However, since the (FRW) Hawking-like 
temperature drops extremely  rapidly as one moves away from the time $t_{rad} ^*$, the baryogenesis will occur 
only in some small time interval  after $t_{rad} ^*$. For example changing the upper limit of integration in \eqref{N1} 
from $\infty$ to $10 t_{rad}^*$ only changes $N_{B-L}$ by about $1\%$ compared to taking the upper limit
as $t = \infty$. Thus although in principle some particle creation always occurs, in practical terms
the particle creation occurs only in a small time range around $t_{rad}^*$. From \eqref{N1} we can obtain $n_B$ which appears in \eqref{eta}; $n_B$ is the absolute value of $N_{B-L}$ divided by the spatial volume of FRW Universe $Volume = 4 \pi c^3 / 3 H^3$. We take the absolute value of $N_{B-L}$ for our purpose since what one calls a baryon or anti-baryon is convention and depends in our case on the sign of the chemical potential $\mu _i$ in equation \eqref{chem-pot} for baryons and anti-baryons. In this way we find that the baryon number density is
\begin{equation}
\label{N2}
n_B = \frac{|N _{B-L}|}{Volume}  = \frac{0.39 H^3 \hbar ^2}{64 \pi ^2 M_* ^2 c^7 (t_{rad} ^*)^2}~.
\end{equation}
In order to obtain $\eta$ as given in \eqref{eta} we now need to find the entropy density $s$. The expression for the entropy density is 
\begin{equation}
\label{entropy}
s = \frac{\rho + p c}{k_B T} \approx \frac{2 \pi^2}{45 (\hbar c )^3 } g_* (k_B T)^3 \rightarrow  
\frac{2 \pi^2}{45 (\hbar c )^3} g_* (k_B T_{FRW})^3 = \frac{H^3 }{180 c^3 \pi } g_* ~.
\end{equation}
The above result is carried out in the same manner used in \eqref{delta-N} to compute the excess of baryons over anti-baryons and under the condition that $\mu \ll k_B T$. To obtain $\rho$ and $p$ one multiplies the 
energy density and pressure respectively for a range of momentum $p$ to $p+dp$ by the Fermi-Dirac distribution 
({\it i.e.} $\frac{1}{e^{(p c \pm \mu)/ k_B T} +1}$) or the Bose-Einstein distribution ({\it i.e.} $\frac{1}{e^{p c/ k_B T} - 1}$) and integrates over $d^3p$. Note for the calculation in \eqref{delta-N} we only considered fermions, since only fermions carry $B-L$ charge in the standard model. However, both bosons and fermions contribute to the entropy density. Also the Bose-Einstein distribution does not have a chemical potential term since once again bosons in the Standard Model do not carry $B-L$ charge and the chemical potential postulated via \eqref{rj} only couples to $B-L$ charge. The expression for $s$ does not in the end even contain the chemical potential $\mu$. This is because the lowest order term in \eqref{entropy} only involves the temperature $T$ since baryons and anti-baryons contribute the same to quantities like $\rho$ and $p$ from \eqref{entropy}, and therefore they contribute the same to the entropy density. In contrast when we calculated the difference between baryon and anti-baryon number via \eqref{delta-N} the leading term involving only $T$ canceled so that there the leading order term was $\mu T^2$ and thus involved the chemical potential. The quantity $g_*$ in \eqref{entropy} is
\begin{equation}
\label{g*}
g_* = \sum_i \frac{7}{8} g_i ^{fermion} +  \sum_j g_j ^{boson} ~,
\end{equation}
where the sums are over the relativistic fermionic and bosonic degrees of freedom. For the Standard Model at the temperatures we are considering one has $g_* \approx 100$. 

Also, in \eqref{entropy} we have taken $k_B T \rightarrow k_B T_{FRW} = \frac{\hbar H}{2 \pi}$, where for our case we have
$H=\frac{1}{2t_{rad}^*}$. The expression for $\eta$ in \eqref{eta} thus follows from \eqref{N2} and \eqref{entropy} and is given by
\begin{equation}
\label{eta2} 
\eta = \frac{n_B}{s} \approx \frac{17.6 \hbar^2 }{16 \pi g_* M_* ^2 c^4 (t_{rad} ^*)^2}
\approx \frac{\hbar ^2}{100 \pi  M_* ^2 c^4 (t_{rad} ^*)^2}~,
\end{equation}
where the reduced Planck mass is $M_* \approx ((\hbar c)/(8 \pi G))^{1/2} \approx 2.4 \times 10 ^{18} \rm {GeV}$. We are now in a position to fix the constant $K_0$ so that we have the desired value of $t_{rad} ^*$ that provides the observed value of $\eta$ from \eqref{eta} for our model.  

To make the statement explicit note that in \eqref{inflation-era} with $H_{(dS)} t < K_0/2$ one is in a pre-inflation stage. Once we have $H_{(dS)} t \sim K_0/2$ the exponential, de Sitter-like expansion of inflationary Universe takes place. Since from \eqref{H} $H_{(dS)}$ is very large the time interval of this FRW Hawking-like driven inflation only needs to last for a time interval of $\Delta t \sim 10^{-44} - 10^{-43}~s$ \cite{modak1, modak2} in order to inflate the scale factor $a(t)$ by the required factor of $\sim 10^{26}$. Since the time during which inflation occurs is so short ({\it i.e.}  $\Delta t \sim 10^{-44} - 10^{-43}~s$) due to the large value of $H_{(dS)}$ the time for the onset of radiation domination, $t_{rad}^*$, is very close to the beginning of inflation, $t_{dS}$ -- in particular $t_{rad}^* = t_{dS} + \Delta t$. 

From \eqref{eta2} we find that in order to obtain the observable $\eta \sim 6 \times 10^{-10}$ we should set $t_{rad}^*\sim 10^{-39}~s$. Thus the onset of the radiation era occurs at a time of about $10^4 t_{Pl}$. Since $\Delta t \sim 10^{-44} - 10^{-43}~s$ this means that $t_{dS} \approx 10^{-39} s$. Using this time in the relationship $H_{(dS)} t \sim K_0/2$ and using $H_{(dS)}$ from \eqref{H} we find that $K_0 \approx 10 ^6$. The constant $K_0$ thus determines the time at which inflation starts and subsequently when baryogenesis begins. Making $K_0$ smaller pushes the time of inflation and baryogenesis closer to the Planck scale while making $K_0$ larger moves it further from the Planck scale. What we find from the above analysis is that the time of the onset of inflation ({\it i.e} $t_{dS} \approx 10^{-39} s$) is about four order of magnitude larger than the Planck time ({\it i.e} $t_{Pl} \sim 10^{-43} s$). Similarly in our model the onset of baryogenesis occurs shortly after our inflation period --  $t_{rad}^* = t_{dS} + \Delta t \approx 10^{-39} s$ since $\Delta t$ is small compared to $t_{dS}$. Thus in this FRW, Hawking-like radiation model both inflation and baryogenesis occur earlier than in the usual picture and the energy scale of inflation and baryogenesis is higher. The idea that
baryogenesis and inflation may be related and occur around the same time was also recently investigated in the works  
\cite{Hertzberg:2013mba} \cite{Hertzberg:2013jba}.

In connection with this last statement we note that the results of the BICEP2 measurement have pointed to a GUT energy scale for inflation of about $10^{16}$ GeV. Thus models that predict higher energy scale for inflationary behavior should be considered as they might be important in light of this observation. Indeed there is a loop quantum gravity inspired model that predicts inflation at the Planck scale \cite{lqg}. Likewise, there is no bound, neither from theory nor observation, that says baryogenesis cannot occur at such a higher energy scale. To check our model further we should check how it fares when compared to other observables once the parameters of the model have been fixed. In our case the only parameter that cannot be determined within our semi-classical approach is $K_0$, which, as we have mentioned parametrizes our ignorance on the energy scale where semi-classical approaches breakdowns and quantum gravity effects become important. Fixing $K_0 \approx 10^6$ sets the transition energy scale where the division between the semi-classical approach and the regime of quantum gravity takes place. Although this value serves the purpose of reproducing the right amount of baryogenesis as needed by observation, this choice of $K_0$ will be much more palatable if it also fits other observations. We shall find that this is indeed the fact with our model. We discuss this in the next section.  

\section{Relevance with the BICEP2/Planck measurements and CMB photons}

The recent measurements based on BICEP2 \cite{bicep2} and Planck \cite{planck} results of CMB B-mode polarization indicate two different values for the tensor-to-scalar modes perturbation ratios, given by $r_{BICEP2} \approx 0.2$ and $r_{Planck} \le 0.11$. Note that
$r_{BICEP2}$ is a specific value while $r_{Planck}$ gives an upper bound. There is an impression that the Planck data is more reliable and probably gives the correct number (there is question about the BICEP2 results -- that some or all of the signal may be due to dust). But as we see below this variation does not affect our conclusion by much. The energy scale of inflation, ${\cal E}_{inf}$, is related to $r$ via the following relationship
\begin{equation}
\label{Einf}
{\cal E}_{inf} = 2.19 \times 10^{16} (r/0.2)^{1/4}~ GeV.
\end{equation}
Since ${\cal E}_{inf}$ depends on the one-fourth power of the ratio $r$, its contribution corresponding to both observations is of the order of unity (if one assumes that $r_{Planck}$ is not too far below its upper value of $0.11$). This approximately gives the energy scale of inflation as ${\cal E}_{inf} \sim 10^{16}~ GeV$, which corresponds to the GUT energy scale. Converting this energy scale to a timescale gives an inflationary time scale of $t_{inf} \sim 10^{-38} - 10^{-39}~s$. This matches the time scale of our model for inflation driven by Hawking-like radiation with the chosen value of $K_0\sim 10^{6}$ which was fixed so as to obtain the observed value of $\eta$ needed to generate the correct amount of baryogenesis in our model.

Another important piece of information about the state of the early Universe is given by the observation of the temperature of the CMB photon at the time of decoupling -- the time when the radiation gas of photons decoupled from the matter. The temperature of CMB photons at the time of decoupling is given by $T_{dec} \sim 3000~K$ and the time of decoupling is given by $t_{dec} \sim 10^5~ {\rm yrs} \approx 10^{12}~{\rm s}$ (this time can be calculated just by equating the radiation and matter energy densities). In our model, with $K_0 \sim 10^6$, we can get an estimate of the temperature of the Universe at the time of decoupling in the following manner. As the FRW Hawking temperature is given by $T \approx \frac{\hbar H}{2 \pi k_B}$, during inflation, where $H \approx t^{-1} \approx 10^{39} s^{-1}$, one finds $T_{inflation} \approx 10 ^{28}~K$. In our model the time when inflation begins and when radiation domination begin are almost the same since the time interval for inflation in our model is very short -- $\Delta t \approx 10^{-44} - 10^{-43}~s$. Thus we can take the temperature at the beginning of the radiation domination era approximately the temperature during inflation, {\it i.e,} $T^* _{rad} \approx T_{inflation} \approx 10 ^{28}~K$. During radiation dominated era the temperature of the photon bath depends on the scale factor as $T \propto \frac{1}{a(t)}$ whereas the relationship between the scale factor and time is $a (t) \propto \sqrt{t}$. We can use these relationships to find out the temperature at decoupling in our model following this simple equation: $T_{dec} = T_{rad}^* \left( \frac{a_*}{a_{dec}} \right) = T_{rad} ^* \sqrt{\frac{t_*}{t_{dec}}} \approx 10^3~K$, where we have used $T^* _{rad} \approx 10 ^{28}~K$, $t_* \approx 10 ^{-39}~s$, $t_{dec} \approx 10 ^{12}~s$. Thus we find that it is possible to use our model of inflation and baryogenesis to reproduce the correct temperature of the CMB photon at decoupling.

\section{Summary and Conclusions}

In this paper we have used the particle creation model of inflation proposed in \cite{modak1, modak2} to address the issue
of baryogenesis which is thought to have occurred soon after the inflation -- in the earliest part of the radiation dominated era. The mechanism for baryogenesis presented here depends on introducing a 
quantum gravity motivated coupling between the space-time and the $B-L$ current of the form given in \eqref{rj}. Such couplings have been considered previously \cite{steinhardt, hook}. The coupling allows one to define a chemical potential as in \eqref{chem-pot} which treats baryons and anti-baryons differently and which is the source of the baryon asymmetry. The final ingredient is that one needs a heat bath with temperature $T$ to drive the production of baryons over anti-baryons as in \eqref{delta-N}. In the baryogenesis model presented here the source of this temperature is the Hawking-like radiation associated with FRW space-time. In a  related work \cite{hook} it was proposed that this thermal bath temperature came of the Hawking radiation of primordial black holes. Here we propose that baryon asymmetry is similarly generated via the Hawking-like radiation of the entire FRW  space-time instead of baryogenesis being seeded by a host of primordial black holes, whose size and number must to fixed to obtain the magnitude of the baryon asymmetry generated. 

The present model of baryogenesis differs from the usual picture of baryogenesis in the sense that it is a (in principle) a never ending process. As long as there is an FRW temperature there will be some baryogenesis via \eqref{delta-N}. However after some early time ({\it i.e.} $t_{rad}^* \approx 10^{-39}~s$) the present mechanism of baryogenesis rapidly shuts off. 
The same comment can be made of the black hole proposal for generating baryon asymmetry -- as long as there are
black holes the baryogenesis mechanism proposed in \cite{hook} will generate excess baryon over anti-baryons. At the present time the black holes present are all astrophysical black holes which have a small Hawking temperature  (in all cases much less than the $\approx 2.7~K$ CMB) and thus a vanishingly small baryon asymmetry generation rate. For the baryogenesis mechanism proposed here as the Universe expands the FRW temperature will drop to the point that the baryogenesis rate will be effectively zero. This is similar to what was found for the ``graceful exit'' from the inflation mechanism in \cite{modak1, modak2} -- as the FRW temperature and particle production rate dropped due to the expansion of the Universe, the de Sitter-like inflationary stage would make transition to a radiation dominated phase. In the present model one has a natural explanation for exiting inflation and for the effective cut-off of baryogenesis -- as the Universe expands the FRW temperature, $T_{FRW}$, rapidly become small. This in turn leads to a transition from de Sitter like expansion to radiation dominated expansion \cite{modak1, modak2} and the effective rate of baryogenesis in the radiation stage also drops rapidly with the decrease in $T_{FRW}$ .

Our model of inflation is now found to be capable to reproduce the observationally measured value for the parameter $\eta$ from equations \eqref{eta}, \eqref{eta2} and without the need for re-heating -- as the Universe expands its temperature smoothly transits from being dominated by the FRW temperature to being dominated by the temperature of the radiation. Our model contains no inflaton field so there is no need for it to decay and re-heat the Universe. The ability to obtain an experimentally acceptable value for $\eta$ comes from choosing the dimensionless integration constant, $K_0$, in \eqref{soln} which sets the time scale for when inflation begins and ends in our model. In order to obtain an acceptable value of $\eta$ we had to choose $K_0 \approx 10 ^6$ which then gave a beginning time of inflation of about $10^{-39}~s$. This is much earlier in time than in the canonical picture of inflation and as well the energy scale at which our model of inflation occurs is higher than in the  canonical picture. This energy scale ${\cal E}_{inf} \approx 10^{16}~ GeV$, however, is roughly the energy scale as determined by the recent BICEP2 results 
\cite{bicep2} and approximately remains so even in light of Planck result \cite{planck}. Furthermore, we are able to reproduce the correct temperature for the CMB photons at the time of decoupling. Our model has some degree of fine-tuning  {\it e.g.}, we need to choose the constants $K_0, D_0$ to take a specific values in order to obtain the observed values of $\eta$ and temperature at decoupling. For $K_0$ this fine-tuning has a simple physical justification (unlike the fine-tuning of the inflaton properties in the standard models of inflation) -- $K_0$ parametrizes our ignorance of the exact scale that divides the quantum gravity regime from the semi-classical domain where we have inflation, baryogenesis, radiation domination and so on. Here it should also be pointed out that using the standard result for the Hawking radiation temperature, as is done in Section 2, to the case with an apparent horizon like the FRW space-time, is not without controversy. The difficulty is related with the global definition of particles with respect to an asymptotically flat space-time which may not be obvious for the FRW space-time. However, even in this case one could simply take the particle creation rate which we use in this work as a phenomenological model which is inspired by Hawking radiation. From this phenomenological standpoint the question then becomes ``Does this give a decent fit with GUT inflation, baryogenesis and CMB photons?"As we have shown above the answer is in the affirmative in all cases. 

Finally, one might also ask whether our ``phenomenological model" detailed above could explain the anisotropies in the CMB spectrum or not. This is an important question and we would like to add a speculative remark in this context. Generally Hawking
radiation is taken to be a pure blackbody spectrum. In light of this it would seem difficult to explain the anisotropies
in the CMB using our model for inflation, since there would then be no initial anisotropies in the initial inflation stage
and as the Universe cooled through later stages and toward decoupling the spectrum would not have an anisotropies. However,
the assumption that Hawking radiation is a pure blackbody is {\it probably} not correct if one take into account the backreaction
of the radiation on the metric. In fact in the tunneling picture of Hawking radiation applied to an FRW space-time, the 
spectrum {\it does} deviate from a pure blackbody due to backreaction \cite{zhu, wilczek}. Thus using the tunneling
picture of Hawking radiation we would speculate that the Hawking radiation which we take as driving our inflationary phase
is not a smooth blackbody spectrum, but has anisotropies due to the backreaction of the radiation on the space-time metric. 
In the post inflationary stage, once photons of the CMB cool down, according to the law discussed at the end is section V (which only depends on the scale factor but not on angular direction) these initial anisotropies, seeded during inflationary stage, should 
still be present. This mechanism for creating anisotropies in the CMB is speculative since it is not clear if the deviations
from a pure blackbody spectrum due to backreaction as detailed in the tunneling picture, would give the correct type of anisotropies
({\it i.e.} correct angular scale, correct magnitude of temperature variation).   

\section{Acknowledgment}
This work is partially supported by the American Physical Society's International Travel Grant Award Program (ITGAP) to DS and SKM. DS thanks ICN-UNAM for the kind hospitality and research support during his visit. SKM is supported by DGAPA fellowship from UNAM.

\end{document}